# A Laboratory-based Hard X-ray Monochromator for High-Resolution X-ray Emission Spectroscopy and X-ray Absorption Near Edge Structure Measurements

G.T. Seidler[*], D.R. Mortensen, A.J. Remesnik, J.I. Pacold, N.A. Ball, N. Barry, M. Styczinski, O.R. Hoidn

Physics Department, University of Washington, Seattle WA 98195-1560

We report the development of a laboratory-based Rowland-circle monochromator that incorporates a low power x-ray (bremsstrahlung) tube source, a spherically-bent crystal analyzer (SBCA), and an energy-resolving solid-state detector. This relatively inexpensive, introductory level instrument achieves 1-eV energy resolution for photon energies of ~5 keV to ~10 keV while also demonstrating a net efficiency previously seen only in laboratory monochromators having much coarser energy resolution. Despite the use of only a compact, air-cooled 10 W x-ray tube, we find count rates for nonresonant x-ray emission spectroscopy (XES) comparable to those achieved at monochromatized spectroscopy beamlines at synchrotron light sources. For x-ray absorption near edge structure (XANES), the monochromatized flux is small (due to the use of a low-powered x-ray generator) but still useful for routine transmission-mode studies of concentrated samples. These results indicate that upgrading to a standard commercial high-power line-focused x-ray tube or rotating anode x-ray generator would result in monochromatized fluxes of order $10^6 - 10^7$ photons/s with no loss in energy resolution. This work establishes core technical capabilities for a rejuvenation of laboratory-based hard x-ray spectroscopies that could have special relevance for contemporary research on catalytic or electrical energy storage systems using transition-metal, lanthanide or noble-metal active species.

[*] Corresponding author: seidler@uw.edu

**I. Introduction**

X-ray absorption fine structure (XAFS) spectroscopy[1-3] and related techniques, such as high-resolution x-ray emission spectroscopy (XES) and nonresonant and resonant inelastic x-ray scattering,[2, 4, 5] have an established and growing importance across multiple fields of science. Historically, these methods both helped motivate and also immensely benefitted from the progressive development of synchrotron x-ray light sources and, most recently, x-ray free electron lasers and table-top ultrafast x-ray sources. While the strength of the scientific case for sources with higher brilliance and finer time resolution is undeniable, it is important to recognize the large body of clearly meritorious, ongoing work using XAFS that does not benefit from such extreme source characteristics. These measurements require only bulk (non-imaging) hard x-ray XAFS that can in many cases even be performed in transmission mode, *i.e.*, thus requiring none of fine focus, high flux, or substantial time resolution. Foremost among these problems in contemporary research are examples in energy sciences, including *in situ* characterization of electrical energy storage[6-27] and of catalysis.[28-48] However, even bulk-like, transmission mode XAFS is exclusively performed at synchrotron light sources; this is not because these studies need the full performance of the beamlines at these facilities, but is instead because of the absence of any alternative.

For the vast majority of other advanced materials and chemical characterization techniques there exists a continuum of instrumentation capabilities having an inverse relationship with their availability: those apparatus with the coarsest performance are inexpensive, widely distributed, and quite easily available for, *e.g.*, teaching purposes or initial sample characterization, while only the absolutely most advanced instruments exist as rare shared-user facilities or unique research instruments. By contrast, and with only rare exceptions in the last 20 years, 'routine' hard x-ray XAFS[49-64] and high-resolution XES[65-70] can only be performed at the synchrotron light sources. The nearly complete restriction of high-resolution hard x-ray spectroscopies to major user facilities is an anomaly in current scientific practice.

The issue here is not just the adverse consequences of finite, infrequent synchrotron access on existing XAFS and XES research programs, but also that those same limitations necessarily exclude a large body of potentially important scientific or industrial work from even being considered. Hence, we propose that the range of applications of XAFS and high-resolution XES has been directly constrained by the limited, infrequent access to synchrotron x-ray facilities, by the technical and financial barriers to implementing the more complex *in situ* studies that are not easily made portable, by proprietary concerns, and also sometimes by system-specific considerations that inhibit sample transport to the light source, *e.g.*, inherent sample fragility, extreme sensitivity to oxidation, or severe biological or radiological safety considerations. We further propose that the range of applications of these methods has been indirectly constrained in a more subtle way: the absence of any 'introductory level' apparatus poses an immense barrier to teaching these methods to the next generation of scientists and hence restricts the scientific diversity of the future synchrotron user community. These observations are not new. Very similar arguments were made in the early era of synchrotron facility development when XAFS beamlines and laboratory-based XAFS system coexisted, when beamline oversubscription was much less severe, and when each of the size and the scientific range of the XAFS community was much smaller.[71]

For the above reasons, we have begun a critical reinvestigation of conventional laboratory x-ray spectroscopies, *i.e.*, those that do not require high beam brilliance or time resolution and that consequently can make use of conventional laboratory x-ray generators such as x-ray sealed tubes and rotating anode sources. Laboratory-based XAFS played an important role in the early development of the technique[71-75] but has seen only sporadic application in the last 20 years.[49, 50, 53, 54, 56-58, 60, 62-64, 76, 77] More specifically, here we report on a research program aimed at developing a true 'introductory level' x-ray absorption near edge structure (XANES) and high-resolution XES capability. Our goal is to develop an instrument that can be relatively inexpensively assembled from existing, commercial low-maintenance components, that requires no special utilities for electricity or instrument cooling, and that is straightforward to calibrate and operate.

Our approach gains some benefit and convenience from the use of current-generation solid-state detectors and the relatively recent development of inexpensive low-power x-ray tubes based on nanotube field emission electron sources,[78, 79] but we find that the largest advantage comes from our use of spherically-bent crystal analyzers (SBCA's) that are now commercially available but that did not exist when laboratory XAFS largely fell out of favor. We find that an SBCA-based scanning monochromator gives quite fine energy resolution while achieving net monochromator efficiencies that were previously found in only the coarsest-resolution laboratory studies of extended XAFS. With this instrument we obtain excellent XANES spectra for concentrated samples in measurement times of several hours and, rather surprisingly, can also obtain synchrotron-quality (bulk-averaged) nonresonant XES in nearly synchrotron-level measurement times. These results establish groundwork for the broad dissemination of these techniques at the desired 'introductory' level, but also establish important performance milestones that can be extrapolated, in the case of XAFS, to the development of a mid-scale facility based on a more powerful conventional source. Such a facility would allow rapid transmission-mode XANES, and transmission-mode extended x-ray absorption fine structure (EXAFS) on useful time scales, while also opening the possibility of high-resolution fluorescence-mode XANES in the laboratory setting.

This manuscript continues as follows. In section II, we survey the prior work in laboratory-based x-ray spectroscopy and, *en route*, motivate the expected, substantial gains that come from the use of SBCA optics in such instruments. In section III we present the design of the monochromator, and its configuration when used for XES or XANES. In section IV, we present and discuss results for both of these techniques, and also provide considerations that encourage the further development and broad application of laboratory-based, high-resolution x-ray spectroscopies. In section V, we conclude.

**II. Prior work in laboratory-based x-ray spectroscopy**
**II. A. X-ray absorption fine structure**

We present in Table I a summary of selected characteristics of many prior conventional laboratory XAFS systems and of the present spectrometer. All prior systems in Table 1 used a Rowland circle monochromator based on cylindrically-bent crystal analyzers (CBCA's) to implement an energy-scanning monochromator. We restrict ourselves to focusing (rather than dispersive) spectrometers that operate in the hard energy range most relevant for energy science applications, i.e., 5-10 keV. Note that Table I is not organized chronologically but is instead ordered from coarsest to finest energy resolution. A strict instrument-to-instrument comparison is made more difficult by the use of different source characteristics, by the different limitations on source operation that followed from the various detectors used, and by the presence of a wide range of analyzer properties such as integral reflectivity, collection solid angle, and simple analyzer curvature (Johann) versus analyzer curvature with surface grinding (Johannson), etc. However, a pragmatic measure of their relative demonstrated efficiencies is given by the monochromatized flux per unit power of their respective x-ray sources, presented in the fourth column of Table I. This efficiency parameter, while imperfect, captures much of the monochromator-to-monochromator variation while also being clearly important for future design consideration, *i.e.*, cost-benefit analysis.

The monochromatized flux per unit generating power in Table I is nonmonotonic but still illustrates one important, general trend in prior work. Although the obtained energy resolutions are always much coarser than any theoretical limit imposed by the integral reflectivity of the analyzer materials, finer energy resolution is typically associated with greatly decreased monochromator efficiency. This effect was well known and had been carefully explained in the associated literature: finer energy resolution when using a CBCA requires collimation out of the Rowland plane with consequent loss in the effective collection solid angle of the monochromator.[71, 75] The exception to the general trend toward poorer efficiency at higher resolution, that of the study of Williams[80], comes mostly from the use of a particularly favorable Johansson-style CBCA having a considerably larger collection solid angle than the optics used in many of the other entries for prior work in Table I. The present study was significantly motivated by the observation that modern, *spherically*-bent crystal analyzers (SBCA's) regularly obtain energy resolution below 1 eV while also having larger collection solid angles than the older CBCA's, especially when compared to CBCA's that have been collimated in the non-dispersive direction to obtain fine energy resolution. This expectation is validated by the final few entries in Table I, which we will present and discuss in Section IV, below.

**II.B. X-ray emission spectroscopy**

The recent history of high-resolution laboratory-based x-ray emission spectroscopy (XES) is rather different than that of XAFS. The information from XES at hard x-ray energies is more restricted and consequently XES has not seen as explosive a growth as synchrotron facilities have proliferated and their brilliance steadily increased. That being said, XES and related techniques do provide crucial information for problems across many fields[5, 81-87], and there has been a steady presence of laboratory-based high-resolution (nonresonant) XES, often using portable instruments that shared time at the synchrotron and in the laboratory. This has been especially true after the development and commercial availability of CCD-based x-ray detectors allowed the highly productive implementation of von Hamos style, wavelength dispersive spectrometers.[65-70] Laboratory-based work using SBCA's is less common, but includes comparative studies of x-ray emission from direct x-ray excitation as opposed to from *K*-capture.[83, 88]

**III. Monochromator Design**

Unlike in prior laboratory XAFS[49, 51, 53-58, 60, 72, 76, 77, 80, 89-98], as we discuss below, our x-ray source and detector are quite compact and low weight. This suggested to us a novel implementation of the Rowland circle geometry that avoids changing the delicate angular-orientation of the SBCA when energy scanning. The general principles of our implementation of the Rowland circle are shown in Fig. 1. Energy scanning is then achieved by symmetrically scanning the source and detector while synchronously making small adjustments to the location of the SBCA so as to track the moving Rowland circle. Schematic representations of XES and XANES measurements using this general approach are presented in Figures 2 and 3, respectively; they are distinguished only by changing the sample position, the x-ray tube orientation, and the definition of what constitutes the 'source' in Fig. 1. The general mechanism and specific implementation of energy scanning is identical in both cases: our instrument is fundamentally a monochromator.

Next, computer-aided design (CAD) renderings of the monochromator, when configured for XES studies, are shown in Figs. 4 and 5; see the figure captions for an initial discussion of the components therein. In the design of any conventional laboratory XAFS instrument, a basic choice must be made concerning which components may, or must, be held stationary and which components must change position or orientation to enable energy scanning. In the common practice[49, 51, 53-55, 57, 58, 60, 72, 76, 77, 80, 89-98] of prior work for conventional laboratory XAFS, the x-ray source was kept stationary as matter of great technical simplification: the high-powered x-ray tubes and rotating anodes that were employed had large mass and required high voltage cables and cooling lines that complicated their motion. This led to the development of the so-called 'linear XAFS spectrometer'.[71, 75] In rare cases where *in situ* studies in extreme environments were needed, the detector was instead kept stationary at fixed orientation and significant effort and engineering creativity was brought to bear on instead scanning the other components.[56]

Here, on the other hand, each of the x-ray tube source and the detector weigh only a few hundred grams; this allows a particularly convenient implementation using just three linear motion stages after an initial tilt-alignment of the SBCA is performed, *i.e.*, the energy scanning model from Fig. 1. Proper alignment of the source and detector with respect to the center of the SBCA is ensured by the 'steering bars' in Figs. 3 and 4. These rectangular cross-section Al-alloy bars pivot immediately at the source and detector (or exit slit) location on the Rowland circle and also at a pin directly underneath the center of the SBCA. Long slots in the steering bars accommodate the changing chord lengths from source or detector to the SBCA upon energy scanning. We emphasize that this configuration is unique in that the orientation of the SBCA in the laboratory frame does not change; all changes in Bragg angle during energy scanning are due to motion of the linear translation stage, resulting in better than 10-μrad resolution and reproducibility in the Bragg angle of the SBCA.

For clarity of presentation, we have omitted a rendering of either our welded He flight path or the radiation enclosure. The He flight path encompasses ~80% of the total linear travel from source to SBCA to detector. The radiation enclosure is fabricated from lead-lined plywood and includes safety interlock switches that are interfaced with the x-ray source controller. The total external dimensions of the radiation enclosure are 170 cm long, 81 cm wide and 61 cm tall. The spectrometer is in the horizontal plane, on the internal floor of the radiation enclosure. Hinged doors on the top face of the radiation enclosure allow access for sample exchange and instrument maintenance. The achievable range of Bragg angles is 74 to 87 degrees and is constrained by the internal dimensions of the radiation enclosure together with the range of motion and placements of the linear stages under the source and detector.

For XES (see Fig. 2), we elaborate that the idealized plane-slab sample has its surface illuminated by the x-ray tube source and rotated by an angle $\phi$ with respect to the line-of-sight to the center of the SBCA; this serves to decrease the apparent angular width in scattering angle $\delta\theta_B$, thus improving energy resolution while retaining a high rate of x-ray fluorescence generation. Photometric calculations based on this geometry are presented in Section IV, below.

With this overview complete, we now present details on the specific components used. While the instrument can be readily reconfigured for different SBCA radii of curvature, it is designed under the assumption that the most common implementation will be for SBCA having a 1-m radius of curvature, i.e., the most commonly available optic. We have used commercial SBCA's (NJ-XRS Tech) and also several made by colleagues at the Advanced Photon Source or in our own lab; all have proven to be acceptable for XANES and XES in our system. To optimize energy resolution and weaken tolerances for component alignment we operate relatively close to backscatter, a condition that usually requires the acquisition or development of a different analyzer for, *e.g.*, each transition metal *K*-edge of interest. This is a logistical, and to some extent financial, disadvantage compared to the older CBCA-based systems which sacrificed energy resolution partially for the convenience of being able to use a single optic over an extremely wide energy range. That being said, a growing variety of SBCA are being fabricated commercially and in numerous research groups, and recent work summarizes multiple possible crystal materials and orientations suitable for each edge or emission line of interest for the hard x-ray regime.[99] Hence the SBCA-availability bottleneck, if it even still exists, is rapidly disappearing. Switching between different SBCAs is not challenging: a laser diode with a weakly diverging beam is placed at the source location and its refocused reflection from the SBCA is used for prealignment. Once the source is replaced and activated we are able to find the optimum SBCA tilt angles and recover the full Rowland geometry in less than one hour.

The x-ray source is a small, air-cooled tube source with an Au anode (Moxtek, Inc.). The source spot size is ~0.4 mm x 0.4 mm and the maximum accelerating potential and electron current are 50 kV and 0.2 mA, respectively, for a peak power of 0.01 kW. By comparison, the tube and rotating anode x-ray generators used in past conventional laboratory XAFS systems typically had 1 kW to as much as 20 kW total power at similar accelerating potentials.[54] The tube source used here has a transmission geometry wherein the Au film anode is deposited on the inner wall of the thin Be vacuum window; this allows the sample to be placed a few mm away from the anode, resulting in a surprisingly high rate of generation of core-holes and hence of resulting x-ray emission, as we discuss in Section IV.

The detector is a silicon drift detector (SDD) having a 25-mm$^2$ active area and integrated signal processing electronics (Amptek, Inc.). The energy resolution of the SDD is safely better than 200 eV, thus strongly rejecting several backgrounds including fluorescence from the instrument shielding, stray scatter, and the analyzer harmonic content of the analyzed signal from the SBCA. This simplifies shielding considerations and allows us to always use the highest accelerating potential and thus highest generating power from the tube source. The signal within a several-hundred eV band surrounding the energy region of interest for a particular study is integrated for each data point. As shown in section IV, the resulting backgrounds are safely below the level of even valence-level XES upon *K*-shell excitation for 3*d* transition metal species.

Repositioning of the detector and source, as needed for energy scanning (see Fig. 1), is accomplished by a pair of dovetail linear stages (Velmex, Inc.) whose ~1-mm pitch lead screws are driven by NEMA-17 stepper motors enabled by integrated controllers (Arcus Technology). The overall approach to tilt-alignment of the SBCA closely follows our group's prior experience with nonresonant inelastic x-ray scattering instrumentation used at synchrotron light sources.[100] The SBCA is mounted to an aluminum plate that is supported by a rod-end bearing and is held by springs against the micrometer tips at the vertical and horizontal tangent points of the optic. SBCA tilt alignment is achieved by high-resolution ball-end micrometers (OptoSigma) that have been modified for stepper motor control. The NEMA-11 stepper motors that drive the micrometers use integrated motor controllers (Arcus Technology). The theoretical resolution of the tilt stage is ~50 nrad per micrometer step. While the achieved accuracy is certainly not at this level, we do find that this system easily enables reproducible, stable alignment to safely better than the expected intrinsic rocking curve widths of ~100 µrad. The tilt stage described above is translated via a heavy-duty crossed-roller bearing linear stage (Parker Motion Control Systems) that is driven by a NEMA-23 motor (Arcus Technologies) having, again, an integral motor controller. Motion control and data collection are performed in the LabView environment using RS-485 and USB communications, respectively.

In Fig. 6, we show $I_{source}(E)$, the spectral intensity of the tube source normalized per unit energy bandwidth and sample solid angle, extrapolated to its maximum power setting. The raw data used for this figure was measured at the lowest tube current of 0.4 µA with the detector 70 cm away to avoid saturation of the SDD. The accelerating potential was 50 kV and the spectrum has been corrected for the energy-dependence of the SDD response and for geometric effects to achieve the desired units of photons/(eV s sr). We estimate 40% systematic uncertainty in the overall level of this spectrum and the relative intensity of the high-energy tail to the main ~10 keV region due to uncertainties in the SDD response function, absorption corrections, and other experimental artifacts. The several sharp emission lines are from elemental x-ray emission from the Au thin-film target and a few weaker Pb lines from the collimators used in this measurement. The low-energy cut-off of the bremsstrahlung spectrum is due to the transmission geometry of the anode.

## IV. Results and Discussion

We present below in Sections IV.A and IV.B results and discussion for XANES and XES using the laboratory monochromator. In each case, we both provide representative results and also detailed discussion of photometrics, both as further explanation of the present instrument design and performance and also as guidelines for future improvements in this type of laboratory-based apparatus.

## IV. A. Laboratory XANES

For XANES, it is useful to begin with photometrics and then present and discuss the experimental results. Unlike nonresonant XES, where a large portion of the broadband flux from the x-ray source is useful, XANES specifically requires excitation in a narrow energy range. Consequently, no laboratory system using a bremsstrahlung source will compete on the basis of flux, brilliance, or any other fine technical metric with the regularly attained performance at hard x-ray synchrotron beamlines. That being said, it is important to note that the flux needed for high-quality XANES measurements is rather modest for the special case of transmission-mode, nonimaging studies of a sample with suitable edge-step magnitude. By means of example, consider the calculations presented in Fig. 7 for the contribution of the 1$s$ shell to the x-ray absorption ($\mu_{1s}$) of Co metal with a modeled thickness of 4 μm, chosen to be the same as the commercially-available reference sample used in the measurement below. For these simulations, we begin with a linear background subtraction from a high-quality transmission-mode study of a Co metal foil from the XAFS Model Compounds database[101] to obtain a model for $\mu_{1s}(E)$. We then continue by simulating the transmission of the indicated *average* number of photons, subject to Poisson statistics, through the sample. The resulting $\mu_{1s}(E)$ for each indicated level of incident average flux per data point is then recovered via Beer's law. As explained below, for the best coordination with present experiment we do not include Poisson statistics for a simulated measurement of the point-by-point incident flux $I_0(E)$. The effects of the exact incident beam spectra on the reference data or on the simulated transmission measurement have not been included.[102] We assess that $10^5$ incident photons per measurement point is publication-quality, but $10^6$ photons per point will be needed to cleanly resolve subtle pre-edge features and higher exposures would be needed as one moves farther away from the absorption edge, into the extended x-ray absorption fine structure (EXAFS) range.

With this context established, we present in Fig. 8 a comparison between the XANES for a 4-μm thick commercial Co metal reference sample (Exafs Materials) measured with the laboratory monochromator and the synchrotron-based Co reference data used above. The SBCA is a commercial Ge(111) analyzer (NJXRS Tech) where we use a few-hundred eV wide band on the SDD to select the (444) harmonic. The tube power was set to its maximum value but the flux was attenuated ~3x by a brass filter to prevent detector saturation from an Au fluorescence line that inconveniently reflects from a higher harmonic of the Ge(111) SBCA in the middle of the Bragg angle range for this study. The resulting attenuated flux on the sample is ~2000 photons/s and the transmitted signal $I_T(E)$ was measured with an integration time of 80 s/point. The detector exit slit was 5 mm, *i.e.*, it provided only some shielding from stray radiation but otherwise gave no particular contribution to the final energy resolution. The incident flux $I_0(E)$ was measured by removing the sample and repeating the same energy scan with an integration time of only 10 s/point; the resulting data set was then fit to a low-order polynomial and that resulting smooth function, *i.e.*, not having Poisson noise, is used for normalization. The absorption coefficient is determined by Beer's law, $\mu(E) = -\ln(I_T(E)/I_0(E))$. Possible systematic errors from irreproducibility between the transmission scan and the $I_0(E)$ scan can be minimized by a future improvement to monitor the overall x-ray tube output flux on a point-by-point basis with any detector outside of the line of sight to the SBCA. That being said, manufacturer specifications for essentially all modern x-ray tube sources indicate fluctuations and drifts in tube power that are well below the level needed to influence our study.

The two XANES spectra in Fig. 8 are in excellent agreement, with only some weak rounding of the mid-edge shoulder at 7712 eV in the laboratory system that is consistent with the consequences of Poisson noise for the incident flux, see Fig. 7. This suggests that the lab monochromator is performing at an energy resolution that is at least not much coarser than the 1.1-eV resolution expected for the double crystal Si (111) monochromator used for the synchrotron reference study.

Returning now to Table I, we have included several entries for the monochromatized flux in our SBCA-based system. The obtained monochromatized fluxes per unit source power are higher, and in most cases much higher, than seen in all of the older CBCA-based instruments.

Based on these results, we anticipate monochromatized fluxes of $10^6$ photons/s to $10^7$ photons/s if a similar apparatus is constructed using a standard 2 kW line-focus x-ray tube source or a high-powered rotating anode source, respectively. As the ~100-μm source size in the dispersive direction for a standard few-kW line-focus x-ray tube is smaller than that for our present low-powered source, the energy resolution will not degrade and may improve; the use of a line-focus (rather than a point focus) leads to some lost efficiency but is not expected to significantly impact energy resolution, even for wider Bragg angle ranges needed to extend the energy range far enough to penetrate at least somewhat into the extended x-ray absorption fine structure (EXAFS) regime.[103] Such an instrument could serve as a mid-scale user facility, capable of rapid screening of large numbers of ambient samples or, *in situ* electrochemical cells in transmission mode and also capable of fluorescence-mode studies for a useful subset of materials synthesis problems. A higher-powered instrument of this type is now under construction by several of the present authors, and the completed instrument will serve as a shared user facility at the University of Washington Clean Energy Institute. Further details will be reported elsewhere after final assembly and commissioning.[103]

**IV.B. Laboratory Nonresonant XES**

In Fig. 9 we present the measured XES from a 1-mm thick CoO pressed powder pellet (powder source: Alfa Aesar, 90%). The sample area facing the source is a 1-cm diameter disk, the total offset from the face-center of the sample to the anode is ~3 mm, and the angle ϕ of the sample face to the SBCA is ~ 15 deg. The SBCA is the same Ge(111) SBCA and same (444) harmonic as used in the XANES study, above, and a few-hundred eV wide acceptance window is again set on the SDD output. The data is collected with the maximum tube output. The integration times are 20 s for the main $K\beta$ energy range and 80 s for the valence region. The data cleanly demonstrate all expected features over the entire energy range, including the weak $K\beta''$ peak due to electron transfer from a ligand semicore shell to the 1$s$ core hole on the Co.[5, 104] We emphasize that no background subtraction has been performed. All stray scatter, shielding fluorescence, harmonic signals from the SBCA, and other backgrounds are efficiently rejected by choosing a ~300-eV wide acceptance window on the SDD. This ability to filter backgrounds using the SDD energy resolution is a considerable instrumental advantage in that it reduces the need for all but the most obvious internal shielding, *i.e.*, immediately around the detector and its electronics.

The demonstrated count rates are impressively high. While raw count rates are seldom reported for concentrated systems in synchrotron XES studies, our prior experience with such measurements at monochromatized synchrotron beamlines finds only 10 times larger count rates for an insertion-device beamline operating at $10^{12}$ photons/s.[105] Consequently, we now move to discuss photometrics for XES. Any effective system for measurement of x-ray emission spectroscopy (XES) must attain a suitable product of fluorescence stimulation rate and net detection efficiency. As the general optical configuration for the lab spectrometer is extremely similar to that used for most XES and other inelastic x-ray scattering methods at synchrotron beamlines,[4, 100, 106-114] we focus here on the useful rate of core-hole creation at the sample by the x-ray tube source. This is the critical parameter that can be used, for example, to predict the relative measurement times for XES with the laboratory spectrometer and with a synchrotron beamline (having known flux and consequent core-hole generation rate).

Consider an x-ray source having a power spectrum $I_{source}(E)$, where $I_{source}(E)$ is normalized by bandwidth and solid angle, *i.e.,* its units are photons/(eV s sr). Assume that this source illuminates a flat sample of thickness $t$ at roughly normal incidence. The rate of core-hole creation in a shell α is then

$$\dot{N}_{core-hole} = \Omega_{sample} \int_0^\infty \frac{I_{source}(E)\,\mu_\alpha(E)}{\mu(E)} (1 - Exp[-\mu(E)t])\, dE, \quad (1)$$

where the sample surface subtends a solid angle $\Omega_{sample}$ as viewed from the source and where $\mu(E)$ is the absorption coefficient from all processes at energy $E$ but $\mu_\alpha(E)$ is the absorption coefficient for only the shell $\alpha$ at energy $E$, *e.g.*, the absorption coefficient for 1*s* ionization of a transition metal species. It is useful to put this into context. A special feature of the low-power tube source used here is the very close approach of samples to the anode, with $\Omega_{sample}$ being as large as 1 sr for favorable cases. Making use of the measured $I_{source}(E)$ (see Fig. 6), we then estimate $\dot{N}_{core-hole} \sim 8 \times 10^{11}$/s for thick, concentrated samples of transition metal species; this estimate has ~40% error due to uncertainties in $I_{source}$ and will be further modified by the exact sample chemistry.

Continuing to the next step in the photometrics, the broadband nature of the excitation from the tube source causes a large fraction of the stimulated x-ray emission to occur so deep in the sample as to be strongly absorbed before escaping. To quantify this issue, we next calculate the rate of x-ray fluorescence escape from the face of the sample in the general direction of the SBCA,

$$\dot{N}_{\alpha\to\beta} = \Omega_{sample} B_{\alpha\to\beta} \int_0^\infty \frac{I_{source}(E)\,\mu_\alpha(E)}{\mu(E)+(\mu_\beta/sin\phi)} \left(1 - Exp[-\left(\mu(E) + \left(\frac{\mu_\beta}{sin\phi}\right)\right)t]\right) dE. \quad (2)$$

In Eq. 2, the emission channel is denoted by $\beta$, the escape angle from the sample toward the SBCA by $\phi$ (see Fig. 2), $\mu_\beta$ is the absorption coefficient of the sample at the energy of the emission channel $\beta$, and $B_{\alpha\to\beta}$ is the branching ratio into emission in channel $\beta$ given that an ionization event has occurred in the shell $\alpha$. The experimentally-useful rate of core-hole generation, $\dot{N}_{\alpha\to\beta}/B_{\alpha\to\beta}$, at $\phi = 15$ deg is then estimated to be $\sim 2 \times 10^{11}$/s for thick, concentrated samples of transition metal species with, again, ~40% errors due to systematic uncertainties in $I_{source}(E)$.

It is important to note that the above estimate is surprisingly large, given that we are using only an inexpensive, very low-power commercial x-ray tube source. By comparison, at the Advanced Photon Source the flux of a typical (monochromatized) XAS beamline falls into the range $\sim 2 \times 10^{10}$/s to $\sim 10^{11}$/s for bending magnet beamlines, depending on upstream concentration optics, and is in the range $\sim 1 \times 10^{12}$/s to $\sim 5 \times 10^{13}$/s for insertion device or wiggler beamlines. These fluxes should be decreased somewhat by sample geometry effects for a strict comparison to the above estimate of $\dot{N}_{\alpha\to\beta}/B_{\alpha\to\beta}$ and alternatively could be increased by a large factor through the less-common use of broadband monochromators, such as is most important when studying very dilute samples.[86] That being said, a key observation remains that is in agreement with the above experimental results: many nonresonant XES studies of concentrated samples can be performed with our laboratory instrument without incurring heroic integration times, and often with measurement times quite comparable to those at synchrotron beamlines.

Finally, given the above results and considerations, it is reasonable to ask whether the use of a more powerful x-ray source could give added benefits for lab-based XES. Somewhat counterintuitively, such an upgrade yields strong advantage in only certain cases. In Table II, we present several relevant characteristics of conventional x-ray generators for the hard x-ray region. While cross-apparatus comparisons are certainly influenced by some fine design details, all such systems can use the same target (anode) materials and typically operate at the same accelerating potentials (~50 kV) and the emitted flux per unit solid angle for each source is therefore roughly proportional to the total electron beam power $P$. For the focusing-geometry monochromator used here, the area of the sample facing the SBCA is independent of the choice of source (it is set by the Bragg angle and the desired energy resolution), and consequently the rate of core-hole generation for any sample is proportional to $P/x^2$, where $x$ is the distance from the anode to the sample. The final two columns in Table II provide this metric for the case where the sample is at the closest approach ($d$) to the anode, as can be achieved for ambient samples, and at closest approach plus 20 mm, as would be needed for samples in special environments such as cryostats, furnaces, or chemical reactors. For ambient samples, $P/d^2$ has only modest benefits when upgrading from the present 10 W tube source to a ~20 kW rotating anode, despite the ~30x higher cost for the more intense source. This is due to the much larger distance from the source anode to the Be vacuum window, a fact mandated by the higher IR heat load on the Be window. For systems in special environments, on the other hand, any substantial offset of the sample itself from the exit window of the x-ray source results in strongly decreased efficiency for the low-power tube source but has less impact on the higher powered sources.

## V. Conclusions

In summary, we report the development and performance of a modern, 'introductory level' instrument for laboratory based studies of x-ray absorption near edge structure (XANES) and high-resolution x-ray emission spectroscopy (XES) in the energy range ~5 keV to ~10 keV. Unlike the earlier generation of Rowland-circle based focusing monochromators for laboratory XAFS, we use a modern spherically-bent crystal analyzer as the Bragg analyzer, thus obtaining much improved efficiency at ~1-eV energy resolution. For transmission-mode XANES of concentrated samples, a few hours measurement time yields relatively quiet data that is in excellent agreement with synchrotron studies. For XES, we find a surprisingly effective system capable of synchrotron-quality results in nearly synchrotron-level measurement times for ambient samples. Given the accessible cost and easy assembly and operations of the present instrument, we believe that this low-performance, high-access approach can have a significant scientific and technical impact in the existing synchrotron x-ray spectroscopy community and also as a low-barrier entry path for new users and their instruction. Furthermore, the demonstrated monochromatic flux and energy resolution serves as proof of principle for the technical viability of a mid-scale XAFS user facility employing a standard few-kW line-focus tube source or higher-powered rotating anode x-ray generator.

**Acknowledgements**

We thank Edward Stern, Tim Elam, John Rehr, and Diego Casa for useful discussions. This work was supported by the U.S. Department of Energy, Basic Energy Sciences under Grant No. DE-FG02-09ER16106 and also by the Office of Science, Fusion Energy Sciences and the National Nuclear Security Administration thought Grant No. DE-SC0008580. This material is also based in part upon work supported by the State of Washington through the University of Washington Clean Energy Institute.

| | Flux in photons/s (energy in $eV$) | Max Power (kW) | Flux/Power (photons/ W s) | Energy resolution (eV) | Mono. Crystal | Analyzer type |
|---|---|---|---|---|---|---|
| Knapp, *et al.*, 1978.[115] | $\sim 1 \times 10^6$ | 1.2 | 800 | 14 | Ge (220) | JS, CBCA |
| Cohen, *et al.*, 1979.[89] | $5 \times 10^7$ (8980) | 12 | 4000 | 10 | LiF (220) | JS, CBCA |
| Georgopoulos and Knapp, 1981.[116] | $4 \times 10^7$ (7000-10000) | 15 | 2700 | 6 | Si 400 | J, CBCA |
| Tohji, *et al.* 1983.[96] | $3 \times 10^5$ (8980) | 12 | 25 | 5-10 | LiF (220) | JS, CBCA |
| Stern, *et al.*, 1980[75] | $5 \times 10^5$ (8980) | 1.2 | 400 | 5 | Si 400 | J, CBCA |
| Yuryev, *et al.*, 2007.[60] | $2.5 \times 10^5$ (8980) | 12 | 20 | 5 | Ge (311) | J, CBCA |
| Yuryev, *et al.*, 2007.[60] | $5 \times 10^4$ (8980) | 12 | 4 | 3 | Ge (311) | J, CBCA |
| Thulke, *et al.*, 1983.[94] | $5 \times 10^4$ (8980) | 12 | 4 | 2 | Si (111) Si (311) Ge (311) | JS, CBCA |
| Williams, 1982.[80] | $3.7 \times 10^5$ (8980) | 2 | 190 | 2 | Ge (333) | JS, CBCA |
| Present | $2 \times 10^4$ (8000) | 0.01 | 2000 | 1 | Si (444) | J, SBCA |
| Present | $8 \times 10^3$ (7100) | 0.01 | 800 | 1 | Ge (620) | J, SBCA |
| Present | $6 \times 10^3$ (7700) | 0.01 | 600 | 1 | Ge (333) | J, SBCA |

**Table I**. A comparison of laboratory XAFS systems using focusing analyzer optics on a Rowland circle, listed in order of improving energy resolution. The abbreviations used for analyzer type are: J = Johann, JS = Johannson, CBCA = cylindrically-bent crystal analyzer, SBCA = spherically-bent crystal analyzer. The overall monochromator efficiency is captured by the ratio of the measured flux to maximum electron beam power in the x-ray generator. Note that the present study achieves high monochromator efficiency at fine energy resolution through the use of SBCA optics.

| X-ray generator | Electron beam power, $P$ (W) | Sample closest approach, $d$ (mm) | $P/d^2$ (W/mm²) | $P/(d+20\text{mm})^2$ (W/mm²) |
|---|---|---|---|---|
| Rotating anode | $20 \times 10^3$ | ~100 | 2.0 | 1.4 |
| High-power tube | $3 \times 10^3$ | 40 | 1.9 | 0.8 |
| Mid-power tube | 300 | 15 | 1.3 | 0.24 |
| Low-power tube | 10 | 3 | 1.1 | 0.02 |

**Table II:** For various laboratory x-ray generators, a comparison of the broadband flux (including both fluorescence lines and bremsstrahlung) per unit sample face area available at the exit window and also recessed by 20 mm, to allow for environmental apparatus. As the target materials and accelerating potentials are similar for all such generators, the total fluxes are roughly proportional to the total electron beam power. See the text for discussion of the $P/d^2$ and $P/(d+20\text{mm})^2$ metrics.

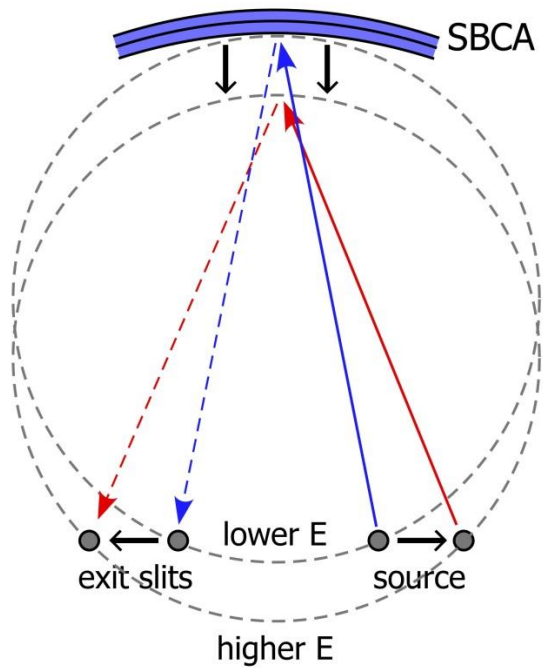

**Figure 1**: Energy scanning of the laboratory x-ray monochromator by synchronized linear motion of the source, the exit slits (and detector), and the spherically-bent crystal analyzer (SBCA). Note the overall symmetry of the configuration and also the simple translation of the Rowland circle.

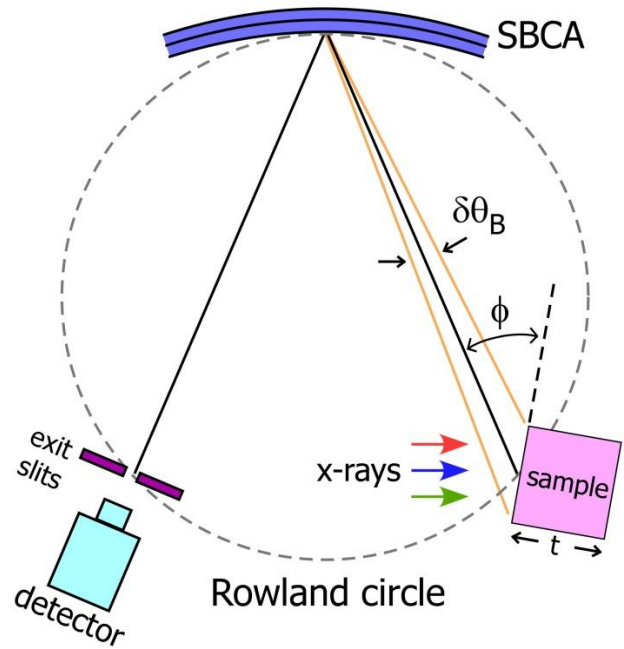

**Figure 2**: The general instrumental configuration for nonresonant x-ray emission spectroscopy with the laboratory monochromator. Broad-band illumination from the x-ray tube source is incident on the face of the idealized sample of thickness *t*. The resulting nonresonant x-ray emission is the analyzed by the spherically-bent crystal analyzer (SBCA) and refocused at the detector. Energy scanning is then implemented as per Fig. 1.

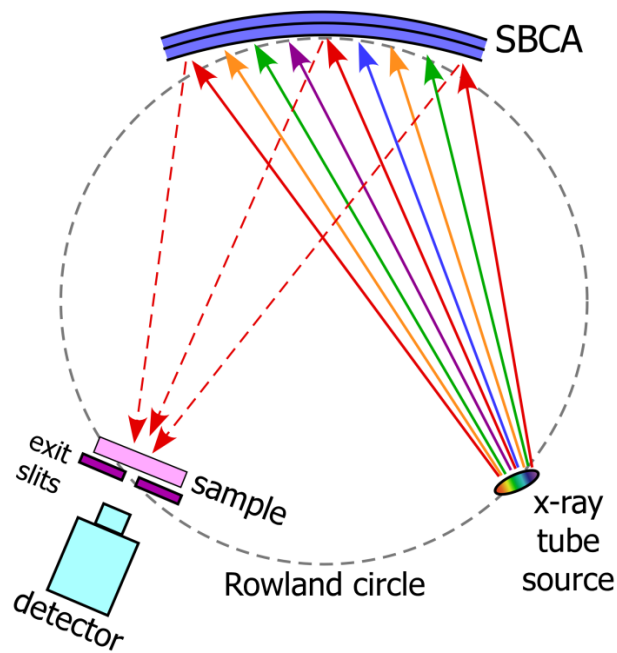

**Figure 3**: The general instrumental configuration for x-ray absorption near edge structure (XANES) studies using the laboratory monochromator. The broadband x-ray radiation from the x-ray tube source directly illuminates the spherically-bent crystal analyzer (SBCA) which monochromatizes and refocuses the radiation onto the sample and the exit slits. The detector measures the transmission through the sample. Energy scanning is then implemented as per Fig. 1. The energy-dependence of the incident flux is characterized by removing the sample from the beampath and repeating the energy scan.

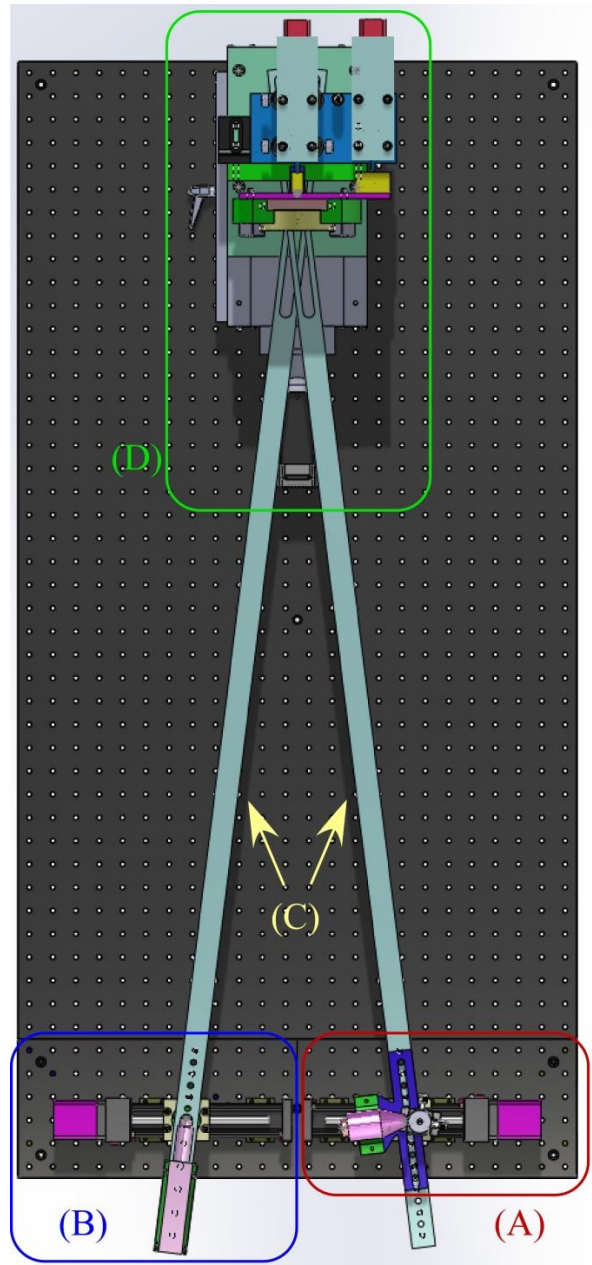

**Figure 4**: Top-view rendering of the Rowland circle monochromator configured for XES measurements. For scale, the spacing of tapped holes in the standard optical breadboard is 25.4 mm. (A): x-ray tube source, sample, manual sample positioner, motorized source-assembly translator; (B): detector, motorized detector translator; (C): steering bars to enforce correct orientation of the source assembly and the detector with respect to the center of the spherically-bent crystal analyzer; (D): two-axis tilt stage, spherically bent crystal analyzer, motorized positioner for linear motion (down the page) of the entire tilt assembly.

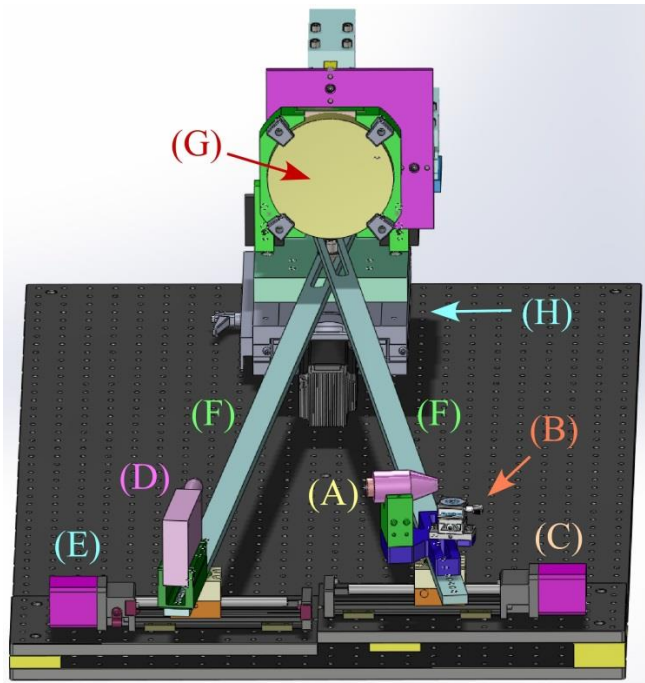

**Figure 5**: Perspective-view rendering of the Rowland circle monochromator configured for XES measurements. For scale, the spacing of tapped holes in the standard optical table is 25.4 mm. (A): x-ray tube source; (B): manual sample positioner; (C): source assembly positioner; (D) detector; (E): detector positioner; (F): steering bars to enforce correct orientation of the source assembly and the detector with respect to the center of the spherically-bent crystal analyzer; (G): spherically-bent crystal analyzer mounted on a two-axis tilt stage; (H): motorized positioner for linear motion (down the page) of the entire tilt assembly.

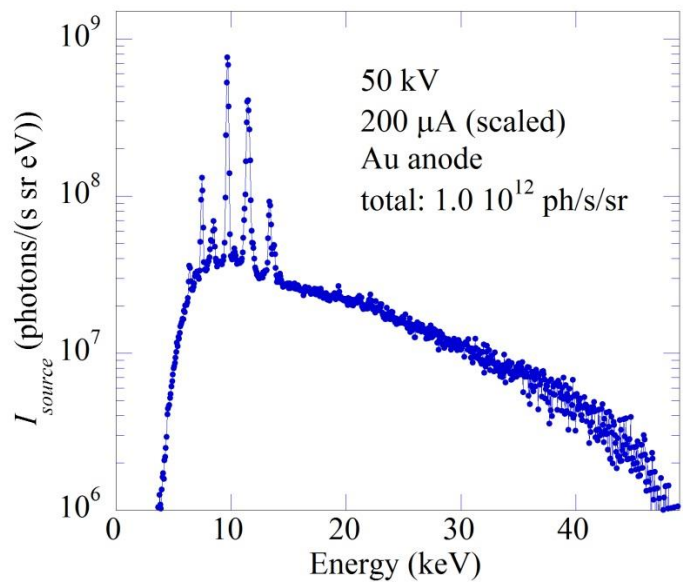

**Figure 6**: The measured spectrum for the x-ray tube source at 50 kV accelerating potential, scaled to the full rated current of 200 µA. Note that the various fluorescence lines are much sharper than shown; the energy resolution of the detector is degraded in this measurement by short shaping times to avoid saturation.

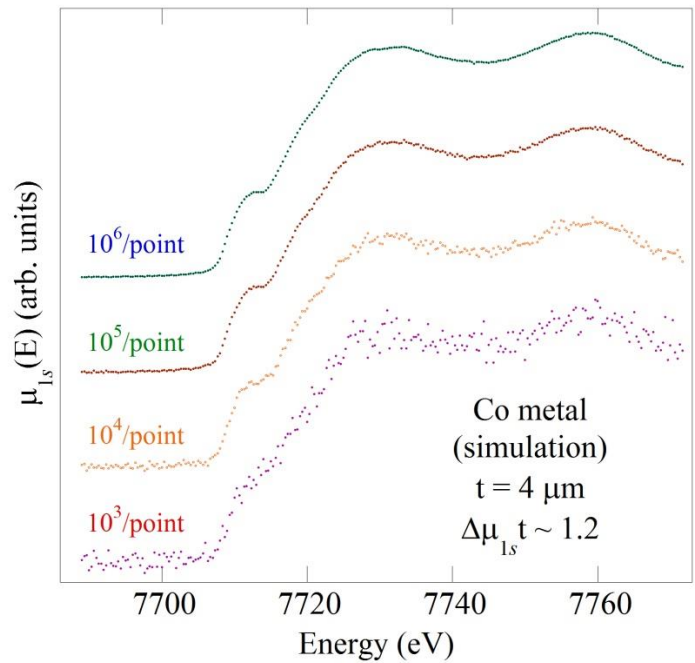

**Figure 7**: Simulated, transmission-mode $\mu_{1s}(E)$ for different numbers of incident photons per data points (as indicated). It is assumed that the Co sample thickness *t* for the simulation is 4 μm so that $\Delta\mu_{1s} \cdot t$ ~1.2 upon crossing the absorption edge. The simulation is based on a Co metal foil reference spectrum that was taken in transmission mode at a synchrotron light source.[101]

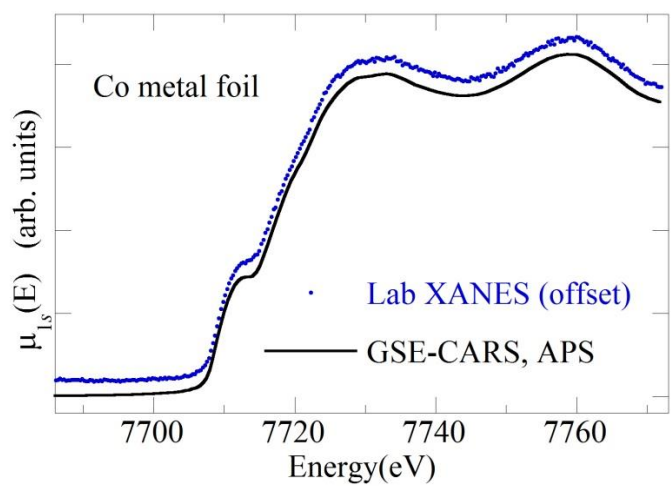

**Figure 8**: XANES for a Co metal foil. For the laboratory XANES data the x-ray tube settings are 50 kV and 200 µA with a brass filter in front of the tube to prevent detector saturation from an Au fluorescence line coincident with a higher Bragg harmonic on the Ge (111) analyzer. Due to the ~3x attenuation of the brass filter, the average incident flux was 2000 photons/s and the integration time for the laboratory XANES data was 80 s per data point. The reference spectrum was taken in transmission mode at a synchrotron light source.[101]

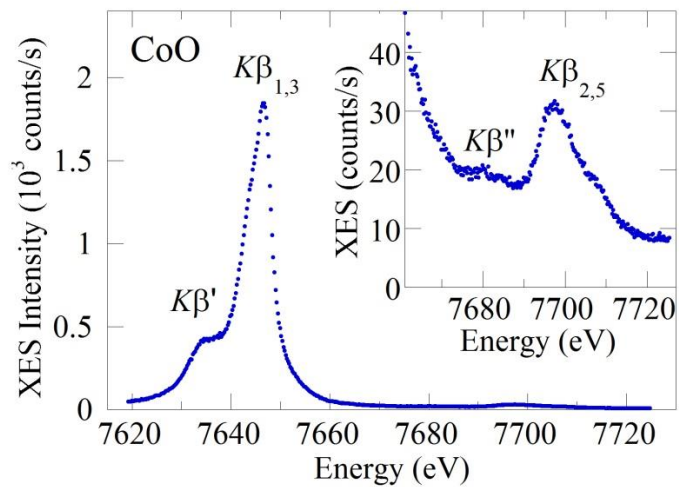

**Figure 9**: Nonresonant XES from a CoO powder sample. The x-ray tube settings were 50 kV and 200 µA. The integration time was 20 s/point in the main $K\beta$ energy range and 80 s/point in the valence region.